\documentclass[table,xcdraw]{article}

\usepackage[english]{babel}
\usepackage[a4paper,top=2cm,bottom=2cm,left=3cm,right=3cm,marginparwidth=1.75cm]{geometry}
\usepackage{amsmath,amsfonts,amssymb,amsthm,mathrsfs}
\usepackage{latexsym,pdfsync,xcolor,graphicx}
\usepackage[colorlinks=true, allcolors=blue]{hyperref}
\usepackage[T1]{fontenc}
\usepackage[utf8]{inputenc}
\usepackage{tikz}
\usepackage{listings}
\usepackage{textcomp}
\usepackage{changepage}
\usepackage{multirow,hhline}

\usepackage[]{algorithm2e}
\usepackage{algpseudocode}
\SetKw{kwpreprocesing}{Preprocesing}
\SetKw{kwspecialization}{Specialization}
\SetKw{kwpowerseries}{Power Series Solution}
\SetKw{kwjacobian}{Jacobian computation}
\SetKw{kwrank}{Rank computation}
\SetKw{kwanalysis}{Analysis}

\newlength{\Oldarrayrulewidth}

\theoremstyle{definition}

\def\R{\ensuremath{\mathbb{R}}}
\def\N{\ensuremath{\mathbb{N}}}
\def\Z{\ensuremath{\mathbb{Z}}}
\def\nei{\ensuremath{\mathcal{N}}}
\def\O{\ensuremath{\mathcal{O}}}
\def\L{\ensuremath{\mathcal{L}}}
\def\K{\ensuremath{\R\left<u\right>(x,\theta)}}

\title{STRIKE-GOLDD 4.0: user-friendly, efficient analysis of structural identifiability and observability}

\author{Sandra Díaz-Seoane$^1$, Xabier Rey Barreiro$^1$, Alejandro F. Villaverde$^{1,2}$\\
{\footnotesize $^1$ Universidade de Vigo, Department of Systems Engineering \& Control, 36310 Vigo, Galicia, Spain}\\
{\footnotesize $^2$ CITMAga, 15782 Santiago de Compostela, Galicia, Spain}
}
\date{\today}

\begin{document}

\maketitle

\abstract{
\noindent 
Structural identifiability and observability are desirable properties of systems biology models. Many software toolboxes have been developed for their analysis in the last decades. STRIKE-GOLDD is a generally applicable tool that can analyse non-linear, non-rational ODE models with unknown inputs. However, this generality comes at the expense of a lower computational efficiency than other tools.
Here we present STRIKE-GOLDD 4.0, which includes a new algorithm, ProbObsTest, specifically designed for the analysis of rational models. ProbObsTest is significantly faster than the FISPO algorithm -- which was already available in older versions of the toolbox -- when applied to computationally expensive models. An important feature of both algorithms is their ability to analyse models with unknown inputs. Thus, their coexistence in the same toolbox provides a combination of general applicability and computational efficiency. 
STRIKE-GOLDD 4.0 is implemented as a free and open-source Matlab toolbox with a user-friendly graphical interface. It is available under a GPLv3 license and it can be downloaded from GitHub at \href{https://github.com/afvillaverde/strike-goldd}{https://github.com/afvillaverde/strike-goldd}.\\
\textbf{Contact:} afvillaverde@uvigo.gal\\
}

\section{Introduction}

A model is structurally identifiable (respectively, observable) if it is theoretically possible to determine its parameters (respectively, state variables) by observations of its output. Structural identifiability and observability (SIO) are properties of dynamic models that are especially important in biological modelling due to the usual experimental limitations \cite{wieland2021structural}.
Hence, in the last two decades many specialized software tools have been developed for their analysis. Structural identifiability can be tested locally or globally. Tools for analysing global and local identifiability include DAISY \cite{bellu2007daisy}, COMBOS \cite{meshkat2014finding}, GenSSI \cite{ligon2018genssi}, SIAN \cite{hong20199sian}, and StructuralIdentifiability \cite{dong2021differential}. Tools that only analyse local identifiability include EAR \cite{karlsson2012efficient}, ObservabilityTest \cite{sedoglavic2002probabilistic}, STRIKE-GOLDD \cite{villaverde2016structural}, and ORC-DF \cite{shi2022efficient}. A critical comparison of their strengths and weaknesses can be found in \cite{rey2022benchmarking}.

\begin{figure*}[ht]
	\includegraphics[width=1.0\linewidth]{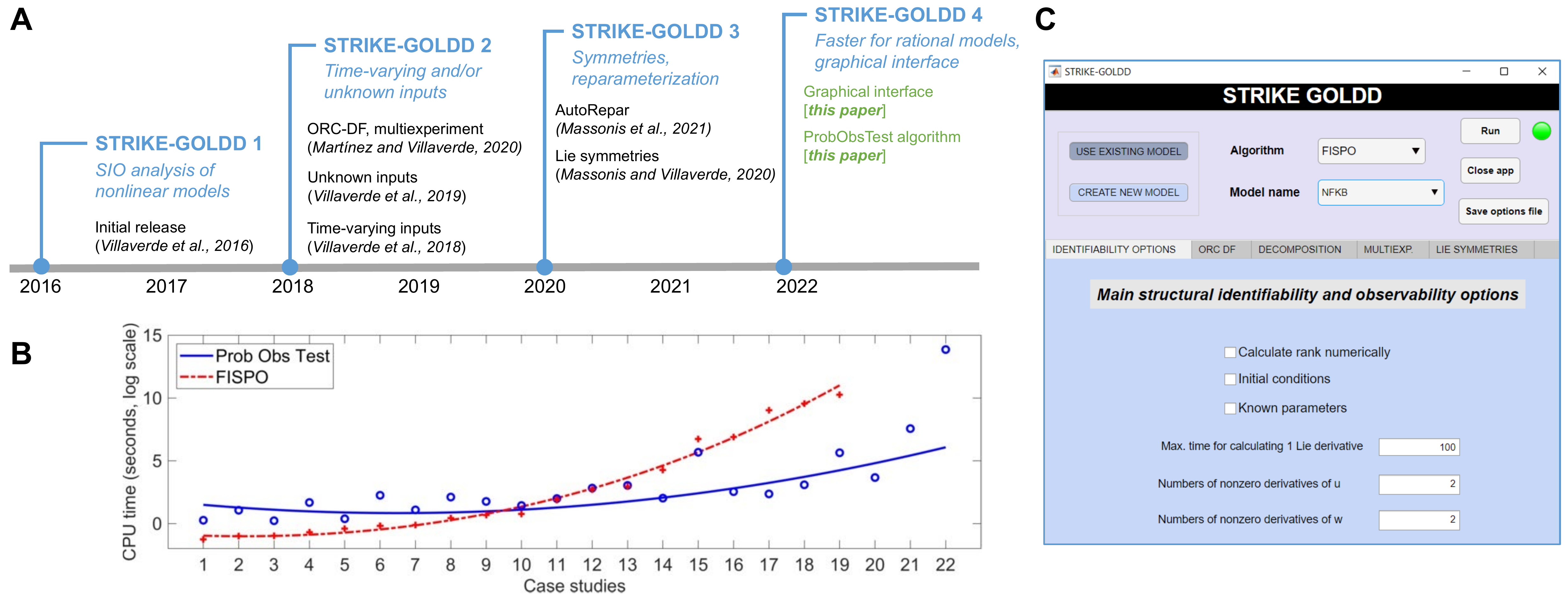}
	\caption{\textbf{(A)} Timeline of the main releases of STRIKE-GOLDD. Each version provides an integrated solution with significantly new features with respect to the previous one: version 2 allowed for an in-depth analysis of the role of inputs, including time-varying and unknown inputs; version 3 allowed searching for the Lie symmetries in the model, and automatically reparameterizing the model to remove said symmetries; version 4 incorporates an algorithm that yields a faster analysis of rational models, as well as a new graphical user interface. These new capabilities are summarized in italics below each main version, along with their publication. 
	\textbf{(B)} Computation times, in seconds, of the new algorithm (Prob Obs Test) vs FISPO algorithm, for 22 case studies of increasing computational complexity. The models and their characteristics are detailed in the Appendix. Circles represent individual computation times, lines are interpolations.
	\textbf{(C)} Graphical interface of the Matlab app STRIKE-GOLDD 4.0.}
	\label{fig:timeline}
\end{figure*}

Most local identifiability tools implement some version of a probabilistic algorithm presented by \cite{sedoglavic2002probabilistic}, which is computationally fast but can only be applied to rational models. In contrast, STRIKE-GOLDD implements an algorithm that is usually less efficient, but allows analysing nonrational models. This versatility is the main strength of STRIKE-GOLDD, which can also analyse models with unknown inputs, search for model symmetries, and find identifiable reparameterizations. A timeline of these developments is shown in Fig. \ref{fig:timeline}.A. However, STRIKE-GOLDD's generality comes at the expense of higher computation times for rational models than other toolboxes tailored to that problem class.  

To address this issue, here we present STRIKE-GOLDD 4.0, which introduces two new features. First, it implements ProbObsTest, an extension of the algorithm by \cite{sedoglavic2002probabilistic} for the analysis of rational models, for which it can achieve considerable speed-ups over FISPO. ProbObsTest can analyse models with unknown inputs, and it includes a procedure for rewriting certain non-rational models in rational form. 
As a second main feature, STRIKE-GOLDD 4.0 is implemented as a Matlab toolbox with a user-friendly graphical interface. 

\section{Methods}

STRIKE-GOLDD 4.0 follows a differential geometry approach \cite{vidyasagar2002nonlinear} to the analysis of structural properties of systems of nonlinear ordinary differential equations (ODE). It analyses the structural local identifiability of their parameters, as well as the observability of their states and unknown inputs. For unidentifiable and/or unobservable models it can search for the Lie symmetries that underlie those deficiencies \cite{massonis2020finding}, and automatically suggest reparameterizations to surmount them \cite{massonis2021autorepar}.

For the first task, SIO analysis, STRIKE-GOLDD 4.0 allows choosing among three algorithms. The most generally applicable is called \underline{FISPO} -- Full Input, State, and Parameter Observability -- which can analyse both non-rational models and models with unknown inputs \cite{villaverde2018input,villaverde2019full}. The FISPO algorithm tests an extended version of the observability rank condition (ORC) \cite{hermann1977nonlinear} that includes unknown inputs as additional states. The ORC test computes an observability matrix from symbolic operations that involve Lie derivatives and calculates its rank. If the rank equals the total number of variables, the model has the FISPO property. The analysis of rational models that are affine in the inputs can also be performed with a second algorithm, \underline{ORC-DF} \cite{martinez2020nonlinear}, which was originally developed by \cite{maes2019observability}.

STRIKE-GOLDD 4.0 includes a third algorithm, \underline{ProbObsTest}, which implements an extension of the algorithm presented in \cite{sedoglavic2002probabilistic} for analysing rational models. This algorithm reduces the computational cost of the ORC test by avoiding the need for symbolic computations of Lie derivatives to build the observability matrix. Instead, it computes a power series expansion whose coefficients are the terms of the observability matrix; since the model variables are specialised on random integer values, the resulting computations as well as the calculation of the rank can be performed numerically. The details of the methodology are provided in the Appendix.

\section{Features and results}

The new algorithm, ProbObsTest, presents two main developments with respect to the algorithm on which it is based, which enlarge the class of models that it can analyse. On the one hand, it has been extended so that it can analyse models with unknown inputs; on the other hand, it can automatically transform certain nonrational models -- such as those with logarithmic and trigonometric functions, or non-integer exponents -- into rational models. These developments are described in the Appendix.

Fig. \ref{fig:timeline}.B. shows a comparison of the computation times of ProbObsTest and FISPO for the analysis of 22 models of increasing complexity.
It can be noticed that for models with a low computational cost (numbers 1--10, which can be analysed in a few seconds or minutes) ProObsTest is slightly slower than FISPO. However, for the more computationally expensive ones (11--22) ProbObsTest is faster, and the difference is larger as the models become more complex. Indeed, while FISPO is unable to analyse models 20--22 due to computational limitations, ProbObsTest performs the calculations successfully. Notably, number 22 is a model of Chinese Hamster Overy cells (CHO) with 117 parameters. Details of the case studies are provided in the Appendix, where we also compare the two algorithms with Sedoglavic's ObservabilityTest for those models to which the latter can be applied.
Additional comparisons of ProbObsTest to other methods can be found in \cite{rey2022benchmarking}.

A second new feature of STRIKE-GOLDD 4.0 is its implementation as a Matlab toolbox with a graphical interface, a screenshot of which is shown in Fig. \ref{fig:timeline}.C. The interface allows choosing the algorithm from a drop-down menu and specifying its settings; likewise, it is possible to select one of the models already existing in the folder from a drop-down menu, or creating a new model from scratch in a new window. This new interface coexists with the previously existing way of executing the toolbox, which is by running a Matlab script; in this case one must indicate the settings by editing an options file. Further details can be found in the user guide included in the documentation folder of STRIKE-GOLDD 4.0.

\section{Conclusion}

STRIKE-GOLDD 4.0 is a user-friendly tool providing a quick way of analysing structural local identifiability and observability. It is arguably the most generally applicable toolbox for local SIO analysis, being able to handle ODE models that may be non-rational and not affine in the inputs, and which may admit unknown external inputs. Previously, this general purpose character came at the expense of high computation times for larger rational models. Version 4.0 addresses this issue by providing a faster algorithm for the analysis of such models, along with the more generally applicable one. The coexistence of both algorithms, as well as the other features included in the toolbox, provides the user with a convenient multi-purpose tool for the analysis and reformulation of systems biology models. Its usability is facilitated further by its implementation as a Matlab app with a user-friendly graphical interface.

\section*{Funding}

This research has received support 
from grant PID2020-113992RA-I00 funded by MCIN/AEI/ 10.13039/501100011033 (PREDYCTBIO);
from grant RYC-2019-027537-I funded by MCIN/AEI/ 10.13039/501100011033 and by ``ESF Investing in your future'';
and from grant ED431F 2021/003 funded by Conseller\'ia de Cultura, Educaci\'on e Ordenaci\'on Universitaria, Xunta de Galicia. 



\section*{APPENDIX}
\appendix

\section{Theory and methods}\label{sec:theory}

\subsection{Background on Structural Identifiability and Observability (SIO)}

Structural identifiability and observability are two key concepts in system identification and dynamic modelling. To give their formal definitions we consider models described by ordinary differential equations with general form:

\begin{equation}\label{mod}
   M:\left\{\begin{aligned}
    \dot{x} & = & f\left(u(t),x(t),\theta\right), \\
    y & = & g\left(u(t),x(t),\theta\right),
\end{aligned}\right. 
\end{equation}
 where $f$ and $g$ are analytic functions (therefore infinitely differentiable); $x(t)\in\R^{n_x}$ is the state variables vector; $u(t)\in\R^{n_u}$, the known inputs vector; $y(t)\in\R^{n_y}$, the outputs vector; $\theta\in\R^{n_\theta}$, the parameters vector. The input vector, $u(t)$, is assumed to consist of infinitely differentiable functions. 
 
A parameter $\theta_i$ of $M$ (\ref{mod}) is \textit{structurally locally identifiable} (s.l.i.) if a neighbourhood $\nei(\theta^*)$ exists such that, for any 
$\hat{\theta}\in\nei(\theta^*)$, $y(t,\theta^*)=y(t,\hat{\theta})$ holds if and only if $\theta^*_i=\hat{\theta}_i$,
for almost any parameter vector $\theta^*\in\R^{n_\theta}$. We say that a parameter is structurally unidentifiable (s.u.) if this relationship does not hold in any $\nei(\theta^*)$. If all the parameters of a model are s.l.i., the model is s.l.i. too. Accordingly, if at least one the parameters is s.u., the model is s.u. 

A s.l.i. parameter can be determined from knowledge of the output $y(t)$ and input $u(t)$ of the model.
Likewise, a state $x_i(\tau)$ is said to be \textit{observable} if it can be determined from the output $y(t)$ and any known inputs $u(t)$ of the model in the interval $t_0 \leq \tau \leq t \leq t_f$, for a finite $t_f$. Otherwise, it is unobservable. A model is observable if all its states are observable, and unobservable if at least one of them is unobservable.

\subsection{SIO analysis with the Observability Rank Condition}

Let us begin with the analysis of observability. The available knowledge for inferring the internal state $x$ of model $M$ consists of the output $y$ and its derivatives. Following a differential geometry approach \cite{dg_an1,dg_an2} we construct a matrix $\O(x)$ that represents a map between the model output $y$ and its derivatives $\dot{y},\ddot{y},...$, on the one hand, and its state $x$ on the other. We can then evaluate the observability by calculating the rank of $\O(x)$. If it has full rank, then the model is observable. If $rank(\O(x_0))=n_x$, $M$ is observable around $x_0$.

With time varying inputs, the output derivatives $\dot{y},\ddot{y},...$ are the so-called ``extended'' Lie derivatives. The extended Lie derivative \cite{karlsson2012efficient} of $g$ with respect to $f$ is defined by:
\[
\L_fg(x)=\frac{\partial g(x)}{\partial x}f(x,u)+\sum_{j=0}^{j=\infty}\frac{\partial g(x)}{\partial u^{(j)}}u^{(j+1)}
\]
and the high order derivatives are recursively calculated:
\[
\L_f^ig(x)=\frac{\partial L_f^{i-1}g(x)}{\partial x}f(x,u)+\sum_{j=0}^{j=\infty}\frac{\partial L_f^{i-1}g(x)}{\partial u^{(j)}}u^{(j+1)}
\]
The $i^{th}$ Lie derivative may contain input derivatives only up to order $(i-1)$ \cite{villaverde2019full}. We can then truncate the infinite summation at $j=i-1$ and rewrite the extended Lie derivative:
\[
\L_fg(x)=\frac{\partial g(x)}{\partial x}f(x,u)+\sum_{j=0}^{j=i-1}\frac{\partial g(x)}{\partial u^{(j)}}u^{(j+1)}
\]
We summarise the way of computing the observability matrix in the following way:
 \begin{equation}\label{obsmat2}
   \O(x(t))=\left(\begin{aligned}
    \frac{\partial}{\partial x}y(t)\;\; \\
    \frac{\partial}{\partial x}\dot{y}(t)\;\;\\
    \vdots\qquad\\
    \frac{\partial}{\partial x}y^{(n_x-1)}(t)
\end{aligned}\right)=\left(\begin{aligned}
    \frac{\partial}{\partial x}g(x)\quad \;\;\\
    \frac{\partial}{\partial x}\left(\L_fg(x)\right)\;\;\\
    \vdots\qquad\quad\\
    \frac{\partial}{\partial x} \left(\L_f^{n_x-1}g(x)\right)
\end{aligned}\right).
\end{equation}

For structural identifiability, parameters can be considered as constant state variables \cite{idasobs1}. Assessing the observability of these states is equivalent to assessing the structural identifiability of the parameters. To this end we construct an augmented state vector $\tilde{x}=\left[x(t);\theta\right]$ with dimension $n_{\tilde{x}}=n_x+n_\theta$ and then we have that $\dot{\tilde{x}}=\left[f(\tilde{x}(t),u(t));\;0\right]$. For a new model with these changes in the variables and equations we construct now an observability-identifiability matrix, $\O_I(\tilde{x}(t))$, in the same way as $\O(x(t))$. Then, if system $M$ given by (\ref{mod}) satisfies $rank(\O_I(\tilde{x}_0))=n_{\tilde{x}}=n_x+n_\theta$, with $\tilde{x}_0$ a point in the augmented state space, the model is observable and identifiable around $\tilde{x}_0$.

\subsection{SIO analysis with the FISPO algorithm in STRIKE-GOLDD}

In \cite{villaverde2016structural} we see that full rank of $\O_I$ might be achieved with less than $n_{\tilde{x}}-1$ Lie derivatives. We also have a minimum number of Lie derivatives, $n_d$, for which the matrix may be full rank. In STRIKE-GOLDD, the $\O_I$ is recursively calculated. Once the $n_d$ Lie derivative is computed, the rank is calculated after adding each new derivative allowing early termination of the procedure. If the full rank is achieved, the OIC is fulfilled; if the rank stops increasing, there is at least one unobservable state or unidentifiable parameter. In the latter case we can determine which parameter or state is unidentifiable or unobservable, respectively. Each column of $\O_I$ corresponds to the partial derivative with respect to a parameter or state. Removing each of the columns and recalculating the rank allows us to know which of the variables is unidentifiable or unobservable. If the rank does not change when we remove the $i^{th}$ column, then the $i^{th}$ variable is unidentifiable or unobservable.

Some models have unmeasured inputs. This is the case of disturbances or time-varying parameters, for example. To analyse them we define a new model that includes them as additional variables:
\begin{equation}\label{mod_w}
   M_w:\left\{\begin{aligned}
    \dot{x} & = & f\left(u(t),w(t),x(t),\theta\right), \\
    y & = & g\left(u(t),w(t),x(t),\theta\right),
\end{aligned}\right. 
\end{equation}
where $w(t)$ refers to the unknown inputs. Then, we define a property analogous to observability for these variables.
An unknown input $w_i(\tau)$ is \textit{reconstructible} if it can be determined from $y(t)$ and $u(t)$ in $t_0 \leq \tau \leq t \leq t_f$, for a finite $t_f$. A model is reconstructible if all its unknown inputs are reconstructible (or ``input observable'').

The property that encompasses observability, structural identifiability, and reconstructibility is called FISPO (full input, state, and parameter observability) \cite{villaverde2019full}. Let $\tilde{x}(t) = [x(t), \theta, w(t)]$ be the vector of unknown model quantities (i.e. states, parameters and inputs), with $z(t)\in\R^{n_x+n_\theta+n_w}$, and
let us denote each element of $\tilde{x}(t)$ at time $\tau$ as $z_i(\tau)$. We say that the model $M$ (\ref{mod}) has the FISPO property if every $\tilde{x}_i(\tau)$ can be determined
from the output $y(t)$ and any known inputs $u(t)$ of the model
in the interval $t_0 \leq \tau \leq t \leq t_f$, for a finite $t_f$. Thus, $M$ is FISPO if,
for every $\tilde{x}_i(\tau)$, for almost any vector $\tilde{x}^*(\tau)$ there is a neighbourhood
$\nei(\tilde{x}^*(\tau))$ in which the following holds:
\[
\hat{\tilde{x}}(\tau) \in \nei(\tilde{x}^*(\tau))\quad and\quad
y(t, \hat{\tilde{x}}(t)) = y(t, \tilde{x}^*(\tau)) )\qquad \Rightarrow \qquad \hat{\tilde{x}}_i(\tau) = \tilde{x}^*_i (\tau).
\]

For assessing reconstructibility, we consider a new augmented state vector 
$\bar{x}=\left[
x(t);
\theta;
w(t)
\right]$
and consequently new state dynamics
$\dot{\bar{x}}=\left[
f(\bar{x}(t),u(t));
0;
\dot{w}(t)
\right]$.
Now, the $i^{th}$ Lie derivative may contain derivatives up to $w^{(i)}$ so, we need to include them in the augmented state vector:
\begin{equation}\label{xfispo}
\bar{x}=\left[\begin{aligned}
x(&t)\\
\theta&\\
w(&t)\\
\dot{w}(&t)\\
\vdots&\\
w^{(i)}&(t)\;
\end{aligned}\right]
\end{equation}
thus the state dynamics are:
\begin{equation}\label{ffispo}
\dot{\bar{x}}=\left[\begin{aligned}
f(\bar{x}(t),&u(t))\\
0\;&\\
\dot{w}(&t)\\
\dot{w}(&t)\\
\vdots\;&\\
w^{(i+1)}&(t)\;
\end{aligned}\right].
\end{equation}
where $\bar{x} \in\R^{n_{\bar{x}}}, n_{\bar{x}}=n_x+n_\theta+n_w(i+1)$.

Let $M_w$ be a model of the form (\ref{mod_w}). $M_w$ is FISPO if the new matrix $\O_I$ computed with the new augmented state vector $\bar{x}(t)$ (\ref{xfispo}) and its corresponding equations $\dot{\bar{x}}(t)$ (\ref{ffispo}) is such that $rank(\O_I(\bar{x},u))=n_{\bar{x}}$. If the matrix is not full rank when it is not computationally feasible or convenient to keep calculating Lie derivatives, the result is inconclusive. To deal with this, we may set to zero the derivatives of $w(t)$ of order higher than a given one $(i)$. We will then have that $w^{(j)}=0,\quad\forall j\geq i$. Even though this assumption restricts the type of inputs that can be analysed, in \cite{villaverde2019full} it was argued by induction that the results might apply to generic inputs under certain circumstances.

\subsection{SIO analysis with a probabilistic algorithm to test local algebraic observability in polynomial time: \textit{ObservabilityTest}}

Sedoglavic presented an algorithm \cite{sedoglavic2002probabilistic} related with the differential algebra approach \cite{da1}, with the goal of computing the set of observable variables of a model in polynomial time. When this technique determines that a variable is observable, the result is guaranteed to be correct. If it classifies it as unobservable, the result is correct with high probability. This approach is applicable to nonlinear rational dynamical systems without unknown inputs. Its definition of algebraic observability is built on the existence of algebraic relations between the state variables and the successive derivatives of the inputs and the outputs. If there is an algebraic relation that allows finitely many trajectories of the state variables that are solutions of the vector field and yield the same specified input-output behavior, the state variables are said to be locally observable \cite{sedoglavic2002probabilistic}. A Maple implementation of this method, called \textit{ObservabilityTest}, is available at \url{https://github.com/sedoglavic/ObservabilityTest/}.

Let us now introduce some notation \cite{sedoglavic2002probabilistic} to formalize the previous definition. We use capital letters to denote the initial conditions of a function and its derivatives, i.e., $u^{(r)}(0)=U^{(r)}$ and $y^{(r)}(0)=Y^{(r)}$ for $r\geq0$ and then $U=(U^{(0)},U^{(1)},...)$, $Y=(Y^{(0)},Y^{(1)},...)$. We denote the field adjoining the indeterminates $U^{(0)}_i,U^{(1)}_i,...$ for $i=1,...,n_u$ and $Y^{(0)}_j,Y^{(1)}_j,...$ for $j=1,...,n_y$ to $\R$ as $\R\left<U,Y\right>$.

We say that $x_i,\;i\in\{1,...,n_x\}$ is locally algebraically observable if $x_i$ is algebraic over the field $\R\left<U,Y\right>$. The system $M$ (\ref{mod}) is locally algebraically observable if the field extension $\R\left<U,Y\right>\hookrightarrow\R\left<U,Y\right>(x)$ is algebraic. The number of non-observable state-variables which should be assumed known, in order to obtain an observable system, can be calculated as the transcendence degree of $\R\left<U,Y\right>\hookrightarrow\R\left<U,Y\right>(x)$. In \cite{sedoglavic2002probabilistic} the transcendence degree is also calculated with the rank of $\O(x(t))$ (\ref{obsmat2}). Thus, if it is a full rank matrix, the transcendence degree is zero and the system is algebraically observable. If it is not full rank, then at least one of the variables is not identifiable and further analysis would be needed to determine which one it is.

While the way of analysing the properties is the same as in the FISPO algorithm, the procedure to compute the matrix and its rank is rather different. In this case, a variational system derived from $M$ (\ref{mod}) is used to directly compute the Jacobian matrix $\O$ \cite{sedoglavic2002probabilistic}; with $x$, $\theta$, and $u$ specialized on some given values. Let us denote by $\Phi(x,\theta,u,t)$ the formal power series in $t$ with coefficients in $\K$ solution of $\dot{\Phi}=f(\Phi,\theta,u)$ with initial condition $\Phi(x,\theta,u,0)=x$. Then:
\[
\Phi(x,\theta,u,t)=x+\sum_{j\in\N^*}\L^jf(x,\theta,u)\frac{t^j}{j!}.
\]
Besides, using $\Phi$, we define the formal power series in $t$ with coefficients in $\K$ as:  
\[
y(x,\theta,u,t)=g(\Phi(x,\theta,u,t),\theta,u,t)=g(x,\theta,u)+\sum_{j\in\N^*}\L^jg(x,\theta,u)\frac{t^j}{j!}
\]
Based on this we have:
\begin{equation}
\begin{aligned}
\O_{alg}=\frac{\partial(y^{(i)})_{0\leq i\leq n_{\tilde{x}}}}{\partial(x,\theta)}&=
\text{coeffs}\left( \frac{\partial g}{\partial x}\frac{\partial \Phi}{\partial x},\frac{\partial g}{\partial x} \frac{\partial \Phi}{\partial \theta} + \frac{\partial g}{\partial \theta} \right)\\
&=\text{coeffs}\left(\nabla y\left( \Phi,\frac{\partial \Phi}{\partial x}, \frac{\partial \Phi}{\partial \theta} \right),j^j,j=0,...,n_{\tilde{x}} \right).
\end{aligned}
\end{equation}
where $\nabla y(\Phi,\Gamma, \Lambda, \theta,u) = \left( \frac{\partial g}{\partial x}\Gamma,\frac{\partial g}{\partial x}\Lambda+\frac{\partial g}{\partial x}\right)(\Phi,\Gamma, \Lambda, \theta,u)$.
Consequently, we have to compute the $n_{\tilde{x}}$ first terms of the power series expansion of $\Phi$, $\Gamma=\frac{\partial\Phi}{\partial x}$ and $\Lambda=\frac{\partial\Phi}{\partial\theta}$.

Since $P(\dot{x},x,\theta,u)=0$, the numerators of the rational relations $\dot{x}-f(x,\theta,u)=0$ and $\nabla P$:
\[
\left\{
\begin{aligned}
&P(\dot{x},x,\theta,u),\\
&\frac{\partial P}{\partial \dot{x}}(x,\theta,u)\dot{\Gamma}+\frac{\partial P}{\partial x}((\dot{x},x,\theta,u)\Gamma,\\
&\frac{\partial P}{\partial \dot{x}}(x,\theta,u)\dot{\Lambda}+\frac{\partial P}{\partial x}((\dot{x},x,\theta,u)\Lambda +\frac{\partial P}{\partial \theta}((\dot{x},x,\theta,u),
\end{aligned}
\right.
\]
the power series $\Phi$,$\Gamma$ and $\Lambda$ are solutions of the system of ordinary differential equations $\nabla P=0$ with the associated initial conditions $\Gamma(x,\theta,u,0)=Id_{n_x\times n_x}$ and $\Lambda(x,\theta,u,0)=0_{n_x\times n_\theta}$.

Next, we specialize the parameters on some random integer $\theta^*$ and the inputs on the power series $u^*$, which are truncated at order $n_{\tilde{x}}+1$ with random integer coefficients. Then, we solve the associated system $\nabla P=0$ for some integer initial conditions $x_0$, and we compute with $\nabla y$ the specialization of $\O_{alg}$.

The Newton operator used in the algorithm is based on the resolution of the following system of linear ordinary differential equations:
\begin{equation}\label{varcte}
\frac{\partial P}{\partial\dot{x}}\dot{E}_{j+1}+\frac{\partial P}{\partial x}E_{j+1}+\nabla P=0\text{  mod  } t^{2j+1},
\end{equation}
with $E_{j+1}=(\Phi-\Phi_j,\Gamma-\Gamma_j,\Lambda-\Lambda_j)$ mod $t^{2j+1}$, correction term,  with $\Phi_j$, $\Gamma_j$ and $\Lambda_j$ are approximations of $\Phi$, $\Gamma$ and $\Lambda$, respectively. The system is solved using $(\Phi_{j+1}$, $\Gamma_{j+1}$, $\Lambda_{j+1})=(\Phi_j$, $\Gamma_j$, $\Lambda_j)+E_{j+1}$ and the initial conditions $\Phi_0\in\Z^{n_x}$, $\Gamma_0=Id_{n_x\times n_x}$ and $\Lambda_0=0_{n_x\times n_\theta}$. The resolution of the linear ordinary differential system relies on the method of integrating factors. We take the homogeneous system
\[
\frac{\partial P}{\partial\dot{x}}(\Phi_j,\theta^*,u^*)\dot{\Omega}_{j}+\frac{\partial P}{\partial x}(\dot{\Phi}_j,\Phi_j,\theta^*,u^*)\Omega_{j}+\nabla P=0\text{  mod  } t^{2j+1},
\]
where $\Omega_{j}$ denotes a $n_x\times n_x$ unknown matrix whose coefficients are truncated series. This homogeneous system is then solved by means of a procedure called ``Homogeneous Resolution'' \cite{sedoglavic2002probabilistic} based on matricial resolution and, in a similar way, the expression in (\ref{varcte}) can be computed with a given precision 
by a procedure called ``Constants Variation'' \cite{sedoglavic2002probabilistic}. Algorithm \ref{algorithm} summarizes the procedure.

\begin{algorithm}
\label{algorithm}
\kwpreprocesing{ Construct a straight-line program encoding the variational system $\nabla P$ and the expressions used during its integration.}

\kwspecialization{Specialization of the parameters, $\theta^*$, and the inputs, $u^*$ }

\kwpowerseries{Computation of the power series solution of $\nabla P$ at order $n_{\tilde{x}}+1$ with a specialised value for the states}

\kwjacobian{Evaluation of $\nabla y$ on the series  $\Phi_j$, $\Gamma_j$ and $\Lambda_j$ where $j=ln_2(n_{\tilde{x}}+1)$, giving the coefficients of the Jacobian matrix}

\kwrank{Calculation of the matrix rank and transcendence degree}

\eIf{transcendence degree = 0} 
    {System is algebraically observable} 
    {Determine which variable or variables are not observable.} 
\caption{Probabilistic algorithm to test local algebraic observability in polynomial time }

\end{algorithm}

\section{New developments in STRIKE-GOLDD 4.0}\label{sec:dev}

\subsection{A new algorithm for SIO analysis in STRIKE-GOLDD: \textit{ProbObsTest}}

STRIKE-GOLDD 4.0 includes a Matlab implementation of an extended version of Sedoglavic's \textit{ObservabilityTest} algorithm. 
The new method is called \textit{ProbObsTest}. It is introduced with the aim of complementing the FISPO algorithm, achieving computational acceleration with respect to it.
In the FISPO algorithm the Lie derivatives are recursively calculated. When their number is relatively small they can be computed in a feasible amount of time. However, for each new Lie derivative that is needed the computational complexity increases, and calculations quickly become impracticable when the number of unknown variables grows. To avoid the need for such calculations, in our implementation of \textit{ProbObsTest} we use the variational system described by Sedoglavic, substituting all the symbolic variables with random numerical values. After this, we get a polynomial matrix that we vectorize, exploiting MATLAB capabilities, in order to obtain an even more efficient implementation. Thus we perform the computations with numerical matrices instead of symbolic expressions. This yields a new code, whose computational complexity increases at a significantly slower rate than FISPO for complex models (as will be demonstrated in Section \ref{sec:comparison}).
The remainder of this section describes the new developments included in \textit{ProbObsTest} with respect to the \textit{ObservabilityTest} algorithm originally presented in \cite{sedoglavic2002probabilistic}.

\subsection{Extending the algorithm to admit unknown inputs}

Sedoglavic's \textit{ObservabilityTest} algorithm cannot analyse models with unknown inputs.
Similarly to the FISPO algorith, \textit{ProbObsTest} can handle unknown inputs by treating them as states. To this end the state is augmented as in (\ref{xfispo}) and the state function $f$ as in (\ref{ffispo}). We can include as many input derivatives as additional states as the number of non-zero derivatives specified as options; the last element of vector $f$ will always be a zero in the presence of unknown inputs. This method cannot handle an infinite number of non-zero derivatives, since the order of the computations grows with the number of derivatives. If the user indicates an infinite number of derivatives in the options, the toolbox automatically lowers it to a relatively small number of non-zero derivatives, warning the user of the change and advising her/him to increase the number if needed, or using the FISPO algorithm to consider the infinite case.

\subsection{Automatic model reformulation: obtaining a rational model}

In principle, \textit{ObservabilityTest} can only be applied to rational models. 
In \textit{ProbObsTest} we have included a procedure to replace non-rational expressions appearing in the model equations with their Taylor expansions. This allows applying the algorithm to non-rational models for which such an expansion is possible; while it is not feasible to do the Taylor expansion for all non-rational functions, this solution covers a significant amount of models. 
The toolbox automatically checks whether it is possible to perform the transformation. If it is, the non-rational expression is replaced. If it is not, the algorithm cannot be applied and an error is issued.

The polynomial numerators of rational terms are obtained with the function $rational2poly-nomial$, which we implemented as an adaptation of the $dagnormal$ procedure in \textit{ObservabilityTest}. Both functions obtain the numerator and denominator of a rational expression. However, we realised that \textit{ObservabilityTest} has problems when non-integer exponents appear in the equations: such terms are non-rational, but the algorithm does not detect it. To fix this issue we automatically approximate such values with the closest integer, which allows applying the procedure while obtaining correct results in the general case.

\subsection{Implementation: a Matlab toolbox with graphical interface}

The  \textit{ProbObsTest} algorithm has been included in STRIKE-GOLDD 4.0, which has been implemented as a Matlab toolbox. A screenshot of its graphical interface is shown in Fig. 1.C of the main text. The interface allows choosing the algorithm from a drop-down menu and specifying its settings; likewise, it is possible to select one of the models already existing in the folder from a drop-down menu, or creating a new model from scratch in a new window. This new interface coexists with the previously existing way of executing the toolbox, which is by running a Matlab script; in this case one must indicate the settings by editing an options file. Further details can be found in the user manual included in the documentation folder of STRIKE-GOLDD 4.0.

\section{Results: comparing FISPO and \textit{ProbObsTest}}\label{sec:comparison}

To determine the computational improvement achieved by \textit{ProbObsTest} with respect to FISPO, we have compared their performance by applying them to a set of 22 problems, which are listed in Table \ref{models}. We considered a number of variations of some of them, so as to assess the effect of varying the number of unknown input derivatives on the performance of the algorithms. The CPU times of both algorithms for the set of 22 models are shown in Figure 1.B of the main text. In the remainder of this section we describe each of the models and discuss the results of the two aforementioned algorithms while also comparing them, when possible, with \textit{ObservabilityTest}.

\begin{table}[ht]
  \centering
  \caption{List of benchmark models and their main features.}
  \label{models}
    \scalebox{1}[1]{
    \begin{tabular}{lcccccc}
    \hline
    \textbf{Model} & \textbf{Ref.}& \textbf{States} & \textbf{Param.} & \begin{tabular}[c]{@{}l@{}}\textbf{Known}\\ \textbf{inputs}\end{tabular} & \begin{tabular}[c]{@{}l@{}}\textbf{Unknown} \textbf{inputs}\\\textbf{(nonzero derivatives)}\end{tabular} & \textbf{Outputs} \\
    \hline
    1: C2M 1& \cite{villaverde2018input,villaverde2019full}& 2&  2&  0&  1(0)&  1 \\
    2: HIV 3& \cite{villaverde2019full,id_nl3}& 3&  5&  1&  0&  2 \\
    3: C2M 2& \cite{villaverde2018input,villaverde2019full}& 2&  4&  1&  0&  1 \\
    4: 2DOF 1& \cite{maes2019observability}& 4&  3&  1&  1(0)& 2  \\
    5: C2M 3& \cite{villaverde2018input,villaverde2019full}& 2&  3&  0&  1(0)&  1  \\
    6: 2DOF 2& \cite{maes2019observability}& 4&  3&  1&  1(2)& 2   \\
    7: C2M 4& \cite{villaverde2018input,villaverde2019full}& 2&  2&  0&  1(3)&  1  \\
    8: 2DOF 3& \cite{maes2019observability}& 4&  3&  0&  2(0)& 2   \\
    9: PK 1& \cite{pk}& 4&  10&  1&  0& 2  \\
    10: C2M 5& \cite{villaverde2018input,villaverde2019full}& 2&  3&  0&  1(3)&  1  \\
    11: PK 2& \cite{pk}& 4&  10&  0&  1(0)& 2  \\
    12: PK 3& \cite{pk}& 4&  10&  0&  1(3)& 2   \\
    13: 2DOF 4& \cite{maes2019observability}& 4&  3&  0&  2(3)& 2   \\
    14: SIRS& \cite{sirs}& 5&  10&  0&  0& 2  \\
    15: NF-$\kappa$B 1& \cite{nfkb,villaverde2016structural}& 15&  13&  1&  0& 6  \\
    16: $\beta$IG 1& \cite{big2}& 3&  5&  1&  0& 1  \\
    17: HIV 5& \cite{hiv5}& 5&  10&  0&  0& 2  \\
    18: $\beta$IG 2& \cite{big2,big1}& 3&  5&  0&  1(0)& 1  \\
    19: NF-$\kappa$B 2& \cite{nfkb,villaverde2016structural}& 15&  29&  1&  0& 6  \\
    20: $\beta$IG 3& \cite{big2}& 3&  5&  0&  1(3)& 1   \\
    21: JAK-STAT& \cite{bjakstat}& 25&  27&  5&  0& 14  \\
    22: CHO& \cite{cho}& 32&  117&  0&  0&  13 \\
    \hline
    \end{tabular}}%
  \label{tab:modelos}%
\end{table}%

\paragraph{C2M model:}
\label{c2m}
The model corresponds to a two compartment system appearing in \cite{villaverde2018input,villaverde2019full}. 
The model consists of:
\begin{itemize}
    \item 2 states, $x=(x_1; x_2)$,
    \item 4 parameters, $\theta=(k_{1e}; k_{12}; k_{21}; b)$,
    \item 1 input, $u$,
    \item 1 output, $x_1$,
\end{itemize}
and it is governed by the following state equations:
\begin{equation*}
    f(x,\theta,u,w)=\left(
    \begin{gathered}
    -(k_{1e}+k_{12})\cdot x_1+k_{21}\cdot x_2+b\cdot u\\
    k_{12}\cdot x_1-k_{21}\cdot x_2
    \end{gathered}
    \right).
\end{equation*}
We start considering the input as known, obtaining a FISPO result from both algorithms. Then, we assume the input as unknown and parameter $b$ as known. Finally, we also assume $k_{1e}$ as known. \textit{ObservabilityTest} can not be applied when there is an unknown input. For the other two methods we also distinguish between constant input and with three non-zero derivatives. Table \ref{tab:c2m} shows the computation times.
\begin{table}[ht]
\selectlanguage{english}
\begin{tabular}{|ll|l|llll|}
\hline
\multicolumn{2}{|l|}{\multirow{2}{*}{Case}}                                                                       & \multirow{2}{*}{\begin{tabular}[c]{@{}l@{}}Known\\ input\end{tabular}} & \multicolumn{4}{l|}{Unknown input}                                                                                                                             \\ \cline{4-7} 
\multicolumn{2}{|l|}{}                                                                                            &                                                                        & \multicolumn{2}{l|}{\begin{tabular}[c]{@{}l@{}}Known\\ $b$\end{tabular}} & \multicolumn{2}{l|}{\begin{tabular}[c]{@{}l@{}}Known\\ $b$ an $k_{1e}$\end{tabular}} \\ \hline
\multicolumn{2}{|l|}{Number of derivatives}                                                                       & \multicolumn{1}{c|}{}                                                  & \multicolumn{1}{c|}{0}              & \multicolumn{1}{c|}{3}             & \multicolumn{1}{c|}{0}                     & \multicolumn{1}{c|}{3}                  \\ \hline
\multicolumn{1}{|l|}{\multirow{3}{*}{\begin{tabular}[c]{@{}l@{}}Execution\\ time\end{tabular}}} & \textit{ObservabilityTest}      & 0,047                                                                  & \multicolumn{1}{l|}{}               & \multicolumn{1}{l|}{}              & \multicolumn{1}{l|}{}                      &                                         \\ \cline{2-7} 
\multicolumn{1}{|l|}{}                                                                          & FISPO    & 0,38                                                                   & \multicolumn{1}{l|}{0,67}           & \multicolumn{1}{l|}{2,14}          & \multicolumn{1}{l|}{0,28}                  & 0,87                                    \\ \cline{2-7} 
\multicolumn{1}{|l|}{}                                                                          & \textit{ProbObsTest} & 1,25                                                                   & \multicolumn{1}{l|}{1,4}            & \multicolumn{1}{l|}{4,19}          & \multicolumn{1}{l|}{1,29}                  & 2,96                                    \\ \hline
\end{tabular}
\caption{Execution times (in seconds) for C2M model}
\label{tab:c2m}
\end{table}

\paragraph{HIV model:}
\label{hiv}
This dynamic system represents a HIV virus infection. We consider two different types of this model. The first one, presented and analysed in \cite{villaverde2019full,id_nl3}, consists of:
\begin{itemize}
    \item 3 states, $x=(T_u; T_I; V)$,
    \item 5 parameters, $\theta=(\lambda; \rho; N; \delta; c)$,
    \item 1 know input, $\eta$,
    \item 2 outputs, $g(x,\theta,w)=(V; T_I + T_u)$,
\end{itemize}
and it is governed by the following state equations:
\begin{equation*}
    f(x,\theta,u,w)=\left(
    \begin{gathered}
    \lambda-\rho\cdot  T_u-\eta \cdot T_u \cdot V\\
    \eta \cdot T_u\cdot  V-\delta\cdot  T_I\\
    N\cdot \delta \cdot T_I-c \cdot V
    \end{gathered}
    \right).
\end{equation*}
The three algorithms yield a FISPO result. \textit{ProbObsTest} takes 2,88 seconds, STRIKE-GOLDD implementation 0,37 and \textit{ObservabilityTest} 0,093.

The other model is introduced in \cite{hiv5}. It has:
\begin{itemize}
    \item 5 states, $x=(xx; y; v; ww; z)$,
    \item 10 parameters, $\theta=(\beta; \lambda; a; b; c; d; hh; k; q; uu)$,
    \item 0 inputs, 
    \item 2 outputs, $g(x,\theta,w)=(ww; z)$,
\end{itemize}
and it is governed by the following state equations:
\begin{equation*}
    f(x,\theta,u,w)=\left(
    \begin{gathered}
    \lambda - (d \cdot  xx) - (\beta \cdot  xx \cdot  v)\\
    (\beta \cdot  xx \cdot  v) - (a \cdot  y)\\
    (k \cdot  y) - (uu \cdot  v)\\
	(c \cdot  z \cdot  y \cdot  ww) - (c \cdot  q \cdot  y \cdot  ww) - (b \cdot  ww)\\
    (c \cdot  q \cdot  y \cdot  ww) - (hh \cdot  z)
    \end{gathered}
    \right).
\end{equation*}
The results obtained are the same for the three algorithms: 3 non observable states and 4 non identifiable parameters. FISPO takes 8237,7  seconds, \textit{ProbObsTest} 10,57 and \textit{ObservabilityTest} 0,203.

\paragraph{2DOF model:}
\label{maes2019observability}
This model is introduced and studied in \cite{maes2019observability}. It is an affine-in-the-inputs model that characterizes the behavior of a mechanical system. 
The model consists of:
\begin{itemize}
    \item 4 states, $x=(x_1; x_2; \dot{x_1}; \dot{x_2})$,
    \item 3 parameters, $\theta=(k_1; \delta k_1; m_2)$,
    \item 1 known input, $u=F_1$,
    \item 1 unknown input, $w=F_2$,
    \item 2 output,
\end{itemize}
and it is governed by the following state equations:
\begin{equation*}
\small
    f(x,\theta,u,w)=\left(
    \begin{gathered}
     \dot{x_1}\\
     \dot{x_2}\\
     (-(k_1+\delta\cdot  k_1\cdot x_1)x_1+k_2\cdot (x_2-x_1)-c_1\cdot \dot{x_1}+c_2\cdot (\dot{x_2}-\dot{x_1})+F_1)/m_1\\
     (k_2\cdot (x_1-x_2)+c_2\cdot (\dot{x_1}-\dot{x_2})+F_2)/m_2
    \end{gathered}
    \right),
\end{equation*}
and the output equations:
\begin{equation*}
    g(x,\theta,w)=\left(
    \begin{gathered}
     x_1\\
    (k_2\cdot (x_1-x_2)+c_2\cdot (\dot{x_1}-\dot{x_2})+F_2)/m_2
    \end{gathered}
    \right).
\end{equation*}
As we have an unknown input, we did the analysis for the constant case and with two non-zero derivatives. In both situations we find that the model is FISPO with both of the algorithms. FISPO spent 0,5 seconds running for the constant case and 0,83 for time-varying one, while \textit{ProbObsTest} required 5,33 and 9,48, respectively.  

Also, to check the behavior with more than one unknown input, we have adapted the model taking the known input as unknown. For the constant case we found that $x_2$ is non observable and $F_1$ and $F_2$ are not reconstructible. The solution obtained matches the one achieved with the previous implementation. FISPO took 0,5 seconds running while the  \textit{ProbObsTest} took 5,33. With two non-zero derivatives we found that $\dot{x_2}$ and $x_2$ are unobservable, and  $F_1$, $\dot{F_1}$, $F_2$ and $\dot{F_2}$ are not reconstructible. The times are now 18,9 for STRIKE-GOLDD, and 20,79 for the new code. \textit{ObservabilityTest} cannot be applied to this model since it has unknown inputs.

\paragraph{Pharmacokinetics model (PK):}
\label{pk}
This example is taken from \cite{pk}. It is a model of the behavior of certain orally administered drugs. It consists of:

\begin{itemize}
    \item 4 states, $x=(x_1; x_2; x_3; x_4 )$,
    \item 10 parameters, $\theta=(k_1; k_2; k_3; k_4; k_5; k_6; k_7; s_2; s_3)$,
    \item 1 input, $u$, that will be considered known and unknown,
    \item 2 outputs, $g(x,\theta,w)=(s_2\cdot x_2; s_3\cdot x_3)$,
\end{itemize}
and it is governed by the following state equations:
\begin{equation*}
    f(x,\theta,u,w)=\left(
    \begin{gathered}
    u_1-(k_1+k_2)\cdot x_1\\
    k_1\cdot x_1-(k_3+k_6+k_7)\cdot x_2+k_5\cdot x_4\\
    k_2\cdot x_1+k_3\cdot x_2-k_4\cdot x_3\\
    k_6\cdot x_2-k_5\cdot x_4
    \end{gathered}
    \right).
\end{equation*}
For the known input case we obtain that $x_2$, $x_3$ and $x_4$ are unobservable and $k_1$, $k_2$, $k_3$, $k_7$, $s_2$ and $s_3$ are unidentifiable. The execution times are 1,95 seconds for STRIKE-GOLDD, 5,81 seconds for \textit{ProbObsTest} and 0,141 seconds for \textit{ObservabilityTest}. When $u$ is an unknown constant input only $k_4$, $k_5$ and $k_6$ are identifiable, all the states are unobservable, and the input is not reconstructible. FISPO takes 6,54 seconds and \textit{ProbObsTest} 7,3. With 3 non-zero derivatives of the input we get the same result; FISPO takes 14,75 seconds and \textit{ProbObsTest} 16,8.

\paragraph{SIRS model:}
\label{sirs}
In \cite{sirs} the transmission of respiratory syncytial virus (RSV) is modeled. We work with the model that takes into account the seasonal nature of transmission through an oscillating contact rate. The population is divided into susceptible ($S$), infected and infectious ($I$), and recovered ($R$) individuals. The model consists of:
\begin{itemize}
    \item 5states, $x=(S; I; R; x_1; x_2)$,
    \item 10 parameters, $\theta=(\nu; b_1; b_0; M; \mu; g)$,
    \item 0 inputs, 
    \item 2 outputs, $g(x,\theta,w)=(I; R)$,
\end{itemize}
and it is governed by the following state equations:
\begin{equation*}
    f(x,\theta,u,w)=\left(
    \begin{gathered}
    \mu - S\cdot \mu - b_0\cdot (1+b_1\cdot x_1)\cdot S\cdot I +  g\cdot R\\
     b_0\cdot (1+b_1\cdot x_1)\cdot S\cdot I-(\nu+\mu)\cdot I\\
	\nu\cdot I - (\mu+g)\cdot R\\
	-M\cdot x_2\\
	M\cdot x_1
    \end{gathered}
    \right).
\end{equation*}
The three implementations find $b_1$ unidentifiable and $x_1$ and $x_2$ unobservable. STRIKE-GOLDD took 70,56 seconds, \textit{ProbObsTest} 7,55 and \textit{ObservabilityTest} 0,125.

\paragraph{\texorpdfstring{NF-$\kappa$B}{Bookmark Version} model:}
\label{nfkb}
NF-$\kappa$B stands for ''nuclear factor kappa-light-chain-enhancer of activated B cells''. We use as reference the analysis done with STRIKE-GOLDD in \cite{villaverde2016structural} based in \cite{nfkb}. The model consists of:
\begin{itemize}
    \item 15 states, $x=(x_1; x_2; x_3; x_4; x_5; x_6; x_7; x_8; x_9; x_{10}; x_{11}; x_{12}; x_{13}; x_{14}; x_{15})$,
    \item 29 parameters, $\theta=(t_1;t_2;c_{3a};c_{4a};c_5;k_1;k_2;k_3;k_{prod};k_{deg};i_1;e_{2a};i_{1a};...\\
    a_1;a_2;a_3;c_{1a};c_{2a};c_{5a};c_{6a};c_1;c_2;c_3;c_4;k_v;e_{1a};c_{1c};c_{2c};c3c)$,
    \item 1 known input, $u_1$, 
    \item 6 outputs: $g(x,\theta,u,w)=\left(
    \begin{gathered}
    x_7\\
    x_{10}+x_{13}\\
    x_9\\
    x_1+x_2+x_3\\
    x_2\\
    x_{12}
    \end{gathered}
    \right)$,
\end{itemize}
and it is governed by the following state equations:
\begin{equation*}
    f(x,\theta,u,w)=\left(
    \begin{gathered}
    k_{prod}-k_{deg}\cdot x_1-k_1\cdot x_1\cdot u_1 \\
    -k_3\cdot x_2-k_{deg}\cdot x_2-a_2\cdot x_2\cdot x_{10}+t_1\cdot x_4-a_3\cdot x_2\cdot x_{13}\dots\\\dots+t_2\cdot x_5+(k_1\cdot x_1-k_2\cdot x_2\cdot x_8)\cdot u_1\\
    k_3\cdot x_2-k_{deg}\cdot x_3+k_2\cdot x_2\cdot x_8\cdot u_1\\
    a_2\cdot x_2\cdot x_{10}-t_1\cdot x_4\\
    a_3\cdot x_2\cdot x_{13}-t_2\cdot x_5\\
    c_{6a}\cdot x_{13}-a_1\cdot x_6\cdot x_{10}+t_2\cdot x_5-i_1\cdot x_6\\
    i_1\cdot k_v\cdot x_6-a_1\cdot x_{11}\cdot x_7\\
    c_4\cdot x_9-c_5\cdot x_8\\
    c_2+c_1\cdot x_7-c_3\cdot x_9\\
    -a_2\cdot x_2\cdot x_{10}-a_1\cdot x_{10}\cdot x_6+c_{4a}\cdot x_{12}\dots\\\dots-c_{5a}\cdot x_{10}-i_{1a}\cdot x_{10}+e_{1a}\cdot x_{11}\\
    -a_1\cdot x_{11}\cdot x_7+i_{1a}\cdot k_v\cdot x_{10}-e_{1a}\cdot k_v\cdot x_{11}\\
    c_{2a}+c_{1a}\cdot x_7-c_{3a}\cdot x_{12}\\
    a_1\cdot x_{10}\cdot x_6-c_{6a}\cdot x_{13}-a_3\cdot x_2\cdot x_{13}+e_{2a}\cdot x_{14}\\
    a_1\cdot x_{11}\cdot x_7-e_{2a}\cdot k_v\cdot x_{14}\\
    c_{2c}+c_{1c}\cdot x_7-c_{3c}\cdot x_{15}
    \end{gathered}
    \right).
\end{equation*}
For this model the execution time with FISPO is 2866,55 seconds while for \textit{ProbObsTest} is 279,51 and for \textit{ObservabilityTest} 8,422 seconds. We found that states $x_8$ and $x_{15}$ are unobservable and $k_2$, $c_4$, $c_{1c}$, $c_{2c}$ and $c_{3c}$ are unidentifiable. We also did the analysis with parameters $a_1$, $a_2$, $a_3$, $c_{1a}$, $c_{2a}$, $c_{5a}$, $c_{6a}$, $c_1$, $c_2$, $c_3$, $c_4$, $k_v$, $e_{1a}$, $c_{1c}$, $c_{2c}$ and $c_{3c}$ fixed to some specific values. In this case we have that $x_{15}$ is unobservable. For the second model the times are 844,46, 291,21 and 3,14, respectively.

\paragraph{\texorpdfstring{$\beta$IG}{Bookmark Version} model:}
\label{big}
This example is one of the four case studies presented in \cite{big2}, representing a physiological circuit that models possible regulatory mechanisms of glucose homeostasis. It was also included in \cite{big1}. 
It consists of:
\begin{itemize}
    \item 3 states, $x=(G; \beta; I)$,
    \item 5 parameters, $\theta=(p; s_i; \gamma; c; \alpha)$,
    \item 1 input, called $u$ that will be considered both as known and unknown,
    \item 1 output, $g(x,\theta,w)=G$,
\end{itemize}
and it is governed by the following state equations:
\begin{equation*}
    f(x,\theta,u,w)=\left(
    \begin{gathered}
     u-(c+s_i\cdot I)\cdot G\\
     \beta\cdot\frac{\frac{0,021}{24\cdot 60}}{1+(\frac{8,4}{G})^{1,7}}-\frac{\frac{0,025}{24\cdot60}}{1+(\frac{G}{4,8})^{8,5}}\\    
     \frac{p\cdot\beta\cdot G^2}{\alpha^2+G^2}-\gamma\cdot I
    \end{gathered}
    \right).
\end{equation*}

With this model we are going to analyse different cases. For all of them the results with the two algorithms match. With the input as known, FISPO takes 984,89 seconds while \textit{ProbObsTest} only 12,65. For constant unknown input these times were 14140,08 and 21,85 seconds, respectively. With three non-zero derivatives of the unknown input, FISPO yields an out of memory error while \textit{ProbObsTest} takes 38,95 seconds. Here we have the best exhibit of the improvements that the new implementation can achieve. For this relatively small but complex system, even for the known input case we see that a great computational acceleration is obtained. Moreover, with the unknown input with three non-zero derivatives we reach a point where FISPO cannot analyse the model while the new algorithm can. 
Finally, in order to analyse this model with \textit{ObservabilityTest} we need to approximate the non-integer exponents in the equations  with integer values. Otherwise, the code issues an error but it gives no clue about its cause or possible fix. After fixing it, the analysis took 0,016 seconds.

\paragraph{JAKSTAT model:}
\label{bjakstat}
This model is introduced in \cite{bjakstat} and consists of:
\begin{itemize}
    \item 25 states, $x=(EpoRJAK2; EpoRpJAK2; p1EpoRpJAK2; p2EpoRpJAK2;\\ p12EpoRpJAK2; EpoRJAK2_CIS; SHP1; SHP1Act; STAT5; pSTAT5;\\ npSTAT5; CISnRNA1; CISnRNA2; CISnRNA3; CISnRNA4; CISnRNA5;\\ CISRNA; CIS; SOCS3nRNA1; SOCS3nRNA2; SOCS3nRNA3\\; SOCS3nRNA4; SOCS3nRNA5; SOCS3RNA; SOCS3)$,
    \item 27 parameters, $\theta=(CISEqc; CISEqcOE; CISInh; CISRNADelay;\\ CISRNATurn; CISTurn; EpoRActJAK2; EpoRCISInh; EpoRCISRemove;\\ JAK2ActEpo; JAK2EpoRDeaSHP1; SHP1ActEpoR; SHP1Dea; SHP1ProOE;\\ SOCS3Eqc; SOCS3EqcOE; SOCS3Inh; SOCS3RNADelay;\\ SOCS3RNATurn; SOCS3Turn; STAT5ActEpoR; STAT5ActJAK2;\\ STAT5Exp; STAT5Imp; init_EpoRJAK2; init_SHP1; init_STAT5)$,
    \item 5 known inputs, $u=(ActD; CISoe; SOCS3oe; SHP1oe; Epo)$,
    \item 0 unknown inputs,
    \item 14 outputs, with equations:
\end{itemize}
\begin{equation*}
    \small
    g(x,\theta,w)=\left(
    \begin{gathered}
    \frac{2\cdot (EpoRpJAK2 + p12EpoRpJAK2 + p1EpoRpJAK2 + p2EpoRpJAK2)}{init_EpoRJAK2}\\
    \frac{16\cdot (p12EpoRpJAK2 + p1EpoRpJAK2 + p2EpoRpJAK2)}{init_EpoRJAK2}\\
    \frac{SOCS3}{SOCS3Eqc}\\
    \frac{STAT5 + pSTAT5}{init_STAT5}\\
    \frac{pSTAT5}{init_STAT5}\\
    STAT5\\
    SHP1 + SHP1Act\\
    CIS\\
    SOCS3\\
    \frac{100\cdot pSTAT5}{STAT5 + pSTAT5}\\
    SOCS3RNA\\
    CISRNA\\
    \frac{(SHP1 + SHP1Act)\cdot (SHP1oe\cdot SHP1ProOE + 1)}{init_SHP1}\\
    \frac{CIS}{CISEqc}
    \end{gathered}
    \right).
\end{equation*} 
For state equations we will refer to \cite{bjakstat_sg}.
To analyse this model we consider the last parameter, which is the initial condition of one of the measured states, as known. 
FISPO yields an out of memory error; to avoid it, the analysis has to be done decomposing the model. In contrast, \textit{ProbObsTest} takes 1914,79 seconds and \textit{ObservabilityTest} just a few seconds.

\paragraph{CHO model:}
\label{cho}
This is a metabolic model of Chinese Hamster Ovary (CHO) cells, which are used for protein production in fermentation processes \cite{cho}. The model consists of:
\begin{itemize}
    \item 32 states, $x=(x_1;\dots; x_{32})$,
    \item 117 parameters, $\theta=(\theta_1; \dots; \theta_{117})$,
    \item 0 inputs,
    \item 13 output, $g(x,\theta,w)=(x_5;x_4;x_3;x_2;x_1;x_{29};x_{27};x_{21};x_{15};x_{13};x_{30};x_{32};x_{11})$.
\end{itemize}
The state equations were modelled with in lin-log kinetics. The resulting expressions are non-rational, due to the presence of logarithms, so \textit{ObservabilityTest} cannot be applied unless the model is reformulated. STRIKE-GOLDD 4.0 checks for this possibility and reformulates the model automatically. 

In \cite{villaverde2016structural}, the analysis of the model with STRIKE-GOLDD (i.e. FISPO) was only partially accomplished. When trying to analyse the whole model FISPO gives an out-of-memory error. After decomposing it in smaller submodels, four parameters were found to be unidentifiable, 95 were classified as identifiable, and the identifiability of the remaining 18 could not be determined. (After fixing six of them, the model was found to be FISPO.) 

In contrast, \textit{ProbObsTest} can analyse the whole model, obtaining conclusive results for all parameters without the need for decomposition. However, it should be noted that it took 12 days to conclude the analysis, which reached the 99\% of the computer memory. 


\begin{thebibliography}{10}

\bibitem{bjakstat}
J.~Bachmann, A.~Raue, M.~Schilling, M.~E. B{\"o}hm, C.~Kreutz, D.~Kaschek,
  H.~Busch, N.~Gretz, W.~D. Lehmann, J.~Timmer, et~al.
\newblock Division of labor by dual feedback regulators controls {JAK2/STAT5}
  signaling over broad ligand range.
\newblock {\em Mol. Syst. Biol.}, 7(1):516, 2011.

\bibitem{bellu2007daisy}
G.~Bellu, M.~P. Saccomani, S.~Audoly, and L.~D’Angi{\`o}.
\newblock {DAISY}: A new software tool to test global identifiability of
  biological and physiological systems.
\newblock {\em Comput. Meth. Prog. Biomed.}, 88(1):52--61,
  2007.

\bibitem{sirs}
M.~A. Capistr{\'a}n, M.~A. Moreles, and B.~Lara.
\newblock Parameter estimation of some epidemic models: the case of recurrent
  epidemics caused by respiratory syncytial virus.
\newblock {\em Bull. Math. Biol.}, 71(8):1890--1901, 2009.

\bibitem{dg_an1}
M.~N. Chatzis, E.~N. Chatzi, and A.~W. Smyth.
\newblock On the observability and identifiability of nonlinear structural and
  mechanical systems.
\newblock {\em Struct. Contr. Health Monitor.}, 22(3):574--593, 2015.

\bibitem{da1}
S.~Diop and M.~Fliess.
\newblock Nonlinear observability, identifiability, and persistent
  trajectories.
\newblock In {\em Proc. 30th IEEE Conference on Decision
  and Control}, pages 714--719. IEEE, 1991.

\bibitem{dong2021differential}
R.~Dong, C.~Goodbrake, H.~A. Harrington, and G.~Pogudin.
\newblock Differential elimination for dynamical models via projections with
  applications to structural identifiability.
\newblock {\em arXiv preprint arXiv:2111.00991}, 2021.

\bibitem{hermann1977nonlinear}
R.~Hermann and A.~Krener.
\newblock Nonlinear controllability and observability.
\newblock {\em IEEE Trans. Autom. Control}, 22(5):728--740, 1977.

\bibitem{hong20199sian}
H.~Hong, A.~Ovchinnikov, G.~Pogudin, and C.~Yap.
\newblock {SIAN}: software for structural identifiability analysis of ode
  models.
\newblock {\em Bioinformatics}, 35(16):2873--2874, 2019.

\bibitem{big2}
O.~Karin, A.~Swisa, B.~Glaser, Y.~Dor, and U.~Alon.
\newblock Dynamical compensation in physiological circuits.
\newblock {\em Mol. Syst. Biol.}, 12(11):886, 2016.

\bibitem{karlsson2012efficient}
J.~Karlsson, M.~Anguelova, and M.~Jirstrand.
\newblock An efficient method for structural identifiability analysis of large
  dynamic systems.
\newblock {\em IFAC proceedings volumes}, 45(16):941--946, 2012.

\bibitem{ligon2018genssi}
T.~S. Ligon, F.~Fr{\"o}hlich, O.~T. Chi{\c{s}}, J.~R. Banga, E.~Balsa-Canto,
  and J.~Hasenauer.
\newblock Genssi 2.0: multi-experiment structural identifiability analysis of
  sbml models.
\newblock {\em Bioinformatics}, 34(8):1421--1423, 2018.

\bibitem{nfkb}
T.~Lipniacki, P.~Paszek, A.~R. Brasier, B.~Luxon, and M.~Kimmel.
\newblock Mathematical model of {NF}-$\kappa${B} regulatory module.
\newblock {\em J. Theor. Biol.}, 228(2):195--215, 2004.

\bibitem{maes2019observability}
K.~Maes, M.~Chatzis, and G.~Lombaert.
\newblock Observability of nonlinear systems with unmeasured inputs.
\newblock {\em Mech. Syst. Signal Process.}, 130:378--394, 2019.

\bibitem{martinez2020nonlinear}
N.~Mart{\'\i}nez and A.~F. Villaverde.
\newblock Nonlinear observability algorithms with known and unknown inputs:
  analysis and implementation.
\newblock {\em Mathematics}, 8(11):1876, 2020.

\bibitem{massonis2021autorepar}
G.~Massonis, J.~R. Banga, and A.~F. Villaverde.
\newblock Autorepar: A method to obtain identifiable and observable
  reparameterizations of dynamic models with mechanistic insights.
\newblock {\em Int. J. Robust Nonlin. Control}, 2021.

\bibitem{massonis2020finding}
G.~Massonis and A.~F. Villaverde.
\newblock Finding and breaking lie symmetries: implications for structural
  identifiability and observability in biological modelling.
\newblock {\em Symmetry}, 12(3):469, 2020.

\bibitem{meshkat2014finding}
N.~Meshkat, C.~E.-z. Kuo, and J.~DiStefano~III.
\newblock On finding and using identifiable parameter combinations in nonlinear
  dynamic systems biology models and {COMBOS}: a novel web implementation.
\newblock {\em PLoS One}, 9(10):e110261, 2014.

\bibitem{id_nl3}
H.~Miao, X.~Xia, A.~S. Perelson, and H.~Wu.
\newblock On identifiability of nonlinear {ODE} models and applications in viral
  dynamics.
\newblock {\em SIAM review}, 53(1):3--39, 2011.

\bibitem{pk}
A.~Raksanyi.
\newblock {\em Utilisation du calcul formel pour l'{\'e}tude des syst{\`e}mes
  d'{\'e}quations polynomiales (applications en mod{\'e}lisation)}.
\newblock PhD thesis, Paris 9, 1986.

\bibitem{rey2022benchmarking}
X.~Rey~Barreiro and A.~F. Villaverde.
\newblock Benchmarking tools for \textit{a priori} identifiability analysis.
\newblock {\em arXiv preprint}, 2022.

\bibitem{sedoglavic2002probabilistic}
A.~Sedoglavic.
\newblock A probabilistic algorithm to test local algebraic observability in
  polynomial time.
\newblock {\em J. Symbol. Comput.}, 33(5):735--755, 2002.

\bibitem{shi2022efficient}
X.~Shi and M.~Chatzis.
\newblock An efficient algorithm to test the observability of rational
  nonlinear systems with unmeasured inputs.
\newblock {\em Mech. Syst. Signal Process.}, 165:108345, 2022.

\bibitem{idasobs1}
E.~Tunali and T.-J. Tarn.
\newblock New results for identifiability of nonlinear systems.
\newblock {\em IEEE Trans. Autom. Control}, 32(2):146--154, 1987.

\bibitem{vidyasagar2002nonlinear}
M.~Vidyasagar.
\newblock {\em Nonlinear systems analysis}.
\newblock SIAM, 2002.

\bibitem{bjakstat_sg}
A.~Villaverde and J.~R. Banga.
\newblock An{\'a}lisis de observabilidad e identificabilidad estructural de
  modelos no lineales: aplicaci{\'o}n a la v{\'\i}a de se{\~n}alizaci{\'o}n
  {JAK/STAT}.
\newblock In {\em XL Jornadas de Autom{\'a}tica}, pages 631--638. Universidade
  da Coru{\~n}a, Servizo de Publicaci{\'o}ns, 2019.

\bibitem{dg_an2}
A.~F. Villaverde.
\newblock Observability and structural identifiability of nonlinear biological
  systems.
\newblock {\em Complexity}, 2019, 2019.

\bibitem{big1}
A.~F. Villaverde and J.~R. Banga.
\newblock Dynamical compensation and structural identifiability of biological
  models: Analysis, implications, and reconciliation.
\newblock {\em PLoS Comput. Biol.}, 13(11), 2017.

\bibitem{villaverde2016structural}
A.~F. Villaverde, A.~Barreiro, and A.~Papachristodoulou.
\newblock Structural identifiability of dynamic systems biology models.
\newblock {\em PLoS Comput. Biol.}, 12(10):e1005153, 2016.

\bibitem{villaverde2018input}
A.~F. Villaverde, N.~D. Evans, M.~J. Chappell, and J.~R. Banga.
\newblock Input-dependent structural identifiability of nonlinear systems.
\newblock {\em IEEE Control Syst. Lett.}, 3(2):272--277, 2018.

\bibitem{cho}
A.~F. Villaverde, D.~Henriques, K.~Smallbone, S.~Bongard, J.~Schmid,
  D.~Cicin-Sain, A.~Crombach, J.~Saez-Rodriguez, K.~Mauch, E.~Balsa-Canto,
  et~al.
\newblock Biopredyn-bench: a suite of benchmark problems for dynamic modelling
  in systems biology.
\newblock {\em BMC Syst. Biol.}, 9(1):1--15, 2015.

\bibitem{villaverde2019full}
A.~F. Villaverde, N.~Tsiantis, and J.~R. Banga.
\newblock Full observability and estimation of unknown inputs, states and
  parameters of nonlinear biological models.
\newblock {\em J. R. Soc. Interface}, 16(156):20190043, 2019.

\bibitem{wieland2021structural}
F.-G. Wieland, A.~L. Hauber, M.~Rosenblatt, C.~T{\"o}nsing, and J.~Timmer.
\newblock On structural and practical identifiability.
\newblock {\em Curr. Opin. Syst. Biol.}, 25:60--69, 2021.

\bibitem{hiv5}
D.~Wodarz and M.~A. Nowak.
\newblock Mathematical models of {HIV} pathogenesis and treatment.
\newblock {\em BioEssays}, 24(12):1178--1187, 2002.

\end{thebibliography}

\end{document}